\documentstyle[twoside,fleqn,npb,epsfig]{article}
\newcommand{\AmS}{{\protect\the\textfont2
  A\kern-.1667em\lower.5ex\hbox{M}\kern-.125emS}}

\newcommand{\be}{\begin{equation}}
\newcommand{\ee}{\end{equation}}
\newcommand{\ba}{\begin{eqnarray}}
\newcommand{\ea}{\end{eqnarray}}

\newcommand{\mc}{\mathcal}

\newcommand{\fr}[2]{{\frac{#1}{#2}\,}}

\renewcommand{\(}{\left(}
\renewcommand{\)}{\right)}

\newcommand{\e}{\epsilon}

\def\sumint{\hbox{$\sum$}\!\!\!\!\!\!\!\int}

\renewcommand{\ln}{{\rm ln}}

\title{QCD pressure up to four loops at finite temperature and density}

\author{A.~Vuorinen\address{Department of Physical Sciences and Helsinki Institute of Physics, \\
        P.O. Box 64, FI-00014 University of Helsinki, Finland}%
        \thanks{aleksi.vuorinen@helsinki.fi}}

\begin{document}

\begin{abstract}
In recent years the perturbative expansion of the pressure of massless QCD has been driven to order $g^6\ln g$ at high temperatures and finite chemical
potentials, which has required calculations up to three-loop order in the full theory and up to four-loop order in three-dimensional effective theories.
In the present paper we briefly review the theoretical background behind this work and explain some of the methods used in the computations.

\end{abstract}

% typeset front matter (including abstract)
\maketitle

\section{INTRODUCTION}
In order to obtain a quantitative picture of the behavior of the QCD pressure in the entire deconfined phase --- not just in the region
$\mu\approx0$, $T\sim T_c$ that is accessible by lattice simulations --- one needs to continuously work to drive its perturbative expansion
to new orders. At high temperatures and zero chemical potential it has been long known that this expansion is plagued by bad convergence properties and
a large renormalization scale dependence, but the highest order ($g^6\ln\,g$) results \cite{klry} derived so far have given reason for a new hope that
the inclusion of the next ${\mc O}(g^6)$ term will improve the situation considerably \cite{lainesewm}. Furthermore, one has been able to show that the
slow convergence is due to the softest bosonic degrees of freedom, which implies that purely fermionic observables, such as quark number
susceptibilities \cite{avsusc,avpres}, behave in a much
cleaner way and that even a simple reorganization of perturbation theory may have significant effects on the properties of the pressure.
Hence there is an obvious motivation to extend the calculations to the next order, where the first fully
non-perturbative contributions from the ultrasoft scale $g^2T$ enter the picture.
%This is the computation I will to some extent outline in the following.

Whereas the high-temperature pressure has been almost constantly under active research during the last three decades, perturbation theory at high
densities and small temperatures has drawn very little attention since the late 1970's. Apart from the emergence of the fundamentally
non-perturbative phenomenon of color superconductivity in this region of the $\mu$-$T$ plane, this is to a large part due to certain computational
problems: having no effective lower-dimensional theories to work with, one is forced to tackle the inevitable IR problems by explicit
resummations of infinite classes of diagrams. This has proven to be a challenging numerical task even at order $g^4$, and it is only very recently
that any progress has been made at small but non-zero temperature \cite{ikrv}.

The purpose of the present report is to introduce to the reader the most important machinery of finite-temperature perturbation theory with special
emphasis on the diagrammatic
tools needed in the determination of the QCD pressure at high $T$. We start by reviewing the concept of dimensional reduction and carry on to describe
the four-loop diagrammatic calculations one encounters both in full QCD and in its three-dimensional effective theories when
trying to extend the perturbative series of the pressure to order $g^6$. To conclude, we then briefly outline some recent efforts to study the behavior
of the low-temperature pressure perturbatively.

\subsection{Basics of finite-$T$ field theory}
The most fundamental quantity that describes the equilibrium thermodynamics of a grand canonical ensemble is its grand potential
\begin{eqnarray}
\Omega(V,T,\mu_i) \;=\; -T\ln\,Z,
\end{eqnarray}
where $Z$ denotes the partition function
\begin{eqnarray}
Z(V,T,\{\mu_i\})\;\equiv\;{\rm Tr\,} {\rm exp}\Big[-\beta\big(H-\sum_i \mu_i N_i\big)\Big],
\end{eqnarray}
$\beta\equiv 1/T$ and the chemical potentials $\mu_i$ correspond to conserved charges. In analogy with the generating functional of the zero-temperature
Green's functions, one can easily derive a functional integral representation for the partition function of an interacting quantum field theory
\begin{eqnarray}
Z\;=\;\int{\mathcal D}\phi \,{\rm exp}\bigg[-\int_{0}^{\beta}{\rm d} \tau \int {\rm d}^3 x \,\mathcal{L}\bigg],
\end{eqnarray}
with $\phi$ being periodic on the interval $[0,\beta]$ for bosons and antiperiodic for fermions. This straightforwardly leads to the finite-temperature
Feynman rules, which differ from the zero-$T$ ones only by the fact that the $p_0$ integrals are replaced by discrete sums
\ba
\int\fr{{\rm d}p_0}{2\pi}&\rightarrow& T \sum_{p_0}
\ea
over the Matsubara frequencies
\begin{eqnarray}
(p_0)_{bos}&=&2\pi nT -i\mu_{bos},\,n\in {\rm Z} \\
(p_0)_{fer}&=&(2n+1)\pi T-i\mu_{fer} , \,n\in {\rm Z}.
\end{eqnarray}
The partition function is finally available through the computation of the one-particle irreducible (1PI) vacuum graphs of the theory.

\section{LARGE $T/\mu$: DIMENSIONAL REDUCTION}
Our eventual goal is to determine the perturbative expansion of the pressure of hot QCD up to order $g^6$ both at zero density and at finite but
moderate quark chemical potentials $\mu_f\leq 10T$, which we in the following will assume to be the case. It has been known for a long time that in this
region the result is not simply a power series in $g^2$ --- which one would naively expect --- but instead contains also odd
powers and logarithms of the coupling constant
\begin{eqnarray}
p_{QCD}\; =\; T^4 \big[A_0 + A_2 g^2 + A_3 g^3 + A_4' g^4\ln\,g \nonumber \\
+ A_4 g^4 + A_5 g^5 + A_6' g^6\ln\,g + A_6 g^6 +
{\mathcal O}(g^7) \big].
\end{eqnarray}
The reason for this is the infrared sensitivity of the quantity, which makes its straightforward diagrammatic expansion diverge already at three loops.
At this order one namely needs to take into account the screening of electrostatic gluons (for which $m_{Debye}\sim gT$) and at four loops also the
non-perturbative screening of the magnetostatic ones ($m_{magn}\sim g^2T$), which implies the necessity to re-organize the perturbative expansion. The
most natural and straightforward means to achieve this is to apply the machinery of dimensional reduction and effective three-dimensional theories, which
in a conceptually simple way leads to the emergence of the terms non-analytic in $g^2$.

In its most straightforward formulation dimensional reduction is simply a statement of the obvious fact that if the temperature is considerably higher
than any other energy scale in the system (in our case most importantly $\Lambda_{QCD}$), the non-static modes of the different fields effectively
decouple because of their large thermal masses. This leaves as the dominant degrees of freedom of high-temperature QCD the electrostatic ($A_0$) and
magnetostatic ($A_i$)
gluons, which, as noted above, however do not remain massless themselves; in order for the setup to work one needs to assume a (parametrically) clear
scale hierarchy from the onset: $g^2T \ll gT \ll 2\pi T$. One may then successively integrate out the scales $2\pi T$ and $gT$, leading to a sequence of
three theories: full QCD describing the hard scales $\sim 2\pi T$, a three-dimensional (because the temporal direction has vanished) Yang-Mills + adjoint
Higgs theory (EQCD) corresponding to the soft ones $\sim gT$, and a 3d pure Yang-Mills theory (MQCD) describing the ultrasoft scales $\sim g^2T$. This
list is exhaustive, since due to confinement there are no more softer scales in the system.

For $\mu=0$ the details of the derivation of the effective theories and their Lagrangians can be found in Refs.~\cite{klry,bn1}, and the minor
modifications that finite chemical potentials induce in Refs.~\cite{avpres,hlp}. The implications the procedure has for the computation of the full
theory pressure are the following: it can be written as a sum of three terms, $p_{QCD}= p_E + p_M + p_G$, that correspond to the scales $2\pi T$, $gT$
and $g^2T$, respectively. Because all the non-perturbativity of the pressure is contained in the last one, the first two can be computed as strict
perturbative expansions, i.e.~pure diagrammatic expansions, in the corresponding theories. Both the IR and UV divergences of these theories can be
handled through dimensional regularization, because in the eventual sum of the three terms they cancel with each other. The function $p_G$, which denotes
the pressure of MQCD and gives its first contributions at orders $g^6\ln\,g$ and $g^6$, has to be computed non-perturbatively, either though a
combination of ordinary lattice simulations and lattice perturbation theory or through stochastic perturbation theory \cite{yorkpar}. It is, however,
notable that the whole contribution of the Linde sea \cite{linde} that for a long time was believed to render the entire expansion of the pressure
useless has now been separated into a well-defined and computable (though yet unknown) number $B_2$
\begin{eqnarray}
p_G\; =\; g^6 \big[ B_1\ln\,g+B_2 \big]+{\mathcal O}(g^8).
\end{eqnarray}

\subsection{The hard scales}

%%%%%%%%%%%%%%%%%%%%%%%%%%%%%%
\begin{figure}[t]
\centerline{\epsfxsize=7.5cm \epsfbox{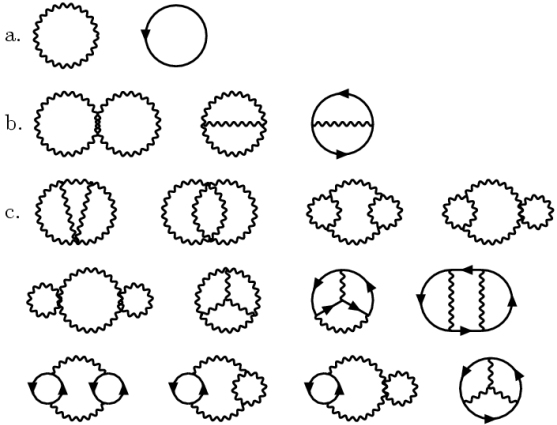}}
\caption[a]{The ghost-free one-, two- and three-loop graphs contributing to $p_E$.}
\end{figure}
%%%%%%%%%%%%%%%%%%%%%%%%%%%%%%%%%%%%

As noted above, the function $p_E$ is computable through the strict perturbation expansion of the full theory pressure, which up to three-loop order (or,
equivalently, $g^6\ln\,g$) is given by the graphs of Fig.~1. All one needs to do in principle is to
write down the expressions of the diagrams, perform the color and Lorentz algebras and finally evaluate the sum-integrals analytically. At the present
order this is indeed enough, as --- at least in the Feynman gauge --- the number of independent sum-integrals that remain to be computed is low enough to
allow for their explicit evaluation by hand. At three loops these `master' sum-integrals are actually all of the type
\begin{eqnarray}
\sumint_{PQR}
\fr{\big(\(P-Q\)^2\big)^m}{P^2Q^2\(R^2\)^{m}\(P-R\)^2\(Q-R\)^2},
\end{eqnarray}
with $P$, $Q$ and $R$ being either bosonic or fermionic and the index $m$ equalling 0, 1 or 2.

In brief, the general strategy in the analytic calculation of sum-integrals of the above type, first developed for vanishing chemical potentials in
Ref.~\cite{az} and later generalized to $\mu\neq 0$ in Refs.~\cite{avpres,antti}, amounts to
\begin{itemize}
\item employing Lorentz invariance and elementary linear changes of integration variables to write the integrands in terms of scalar
`polarization' functions of the type
\ba
\Pi(P)\;\equiv\;\sumint_Q \fr{1}{Q^2(Q-P)^2},
\ea
where no spatial components of the momenta appear in the numerators,
\item subtracting the leading UV parts of the polarization functions from the sum-integrals and treating the corresponding divergent terms analytically
in momentum space,
\item setting $\e\! =\! 0$ in the convergent parts and performing a three-dimensional Fourier transform into coordinate space,
\item performing the $p_0$ sums and in the end analytically solving the remaining (hyperbolic) coordinate space integrals.
\end{itemize}
The results for the purely bosonic sum-integrals are expressible in terms of a few mathematical constants, such as rationals, $\ln\,2$, the Euler
constant $\gamma$, and the derivatives of the Riemann zeta function $\zeta'(-1)$ and $\zeta'(-3)$. For the fermionic cases at finite chemical potential
one in addition runs into the special functions $\zeta'(x,y)\equiv \partial_x \zeta(x,y)$ and $\psi(x)\equiv\Gamma'(x)/\Gamma(x)$ that occur in the
combinations
\ba
\aleph(n,z) &\equiv& \zeta'(-n,z)+\(-1\)^{n+1}\zeta'(-n,z^{*}), \\
\aleph(z) &\equiv& \Psi(z)+\Psi(z^*),
\ea
where $z\equiv 1/2-i\mu/(2\pi T)$ and $n\in\{0,1,2,3\}$. For a somewhat detailed account of the asymptotic properties of these functions, see
Ref.~\cite{avdis}.

Proceeding further to four loops --- one of the tasks required when approaching order $g^6$ in the expansion of the pressure --- the computations become
considerably more involved. The sheer number of
diagrams becomes so large that the automatization of the color and tensor algebras becomes imperative, and one in addition needs a more systematic and
less time-consuming method for dealing with the remaining sum-integrals. This time the number of naive `masters' will easily be ${\mc O}(100)$, which
one definitely needs to reduce in order for an analytic treatment to be possible. The problem, however, is that most of the conventional methods for
finding linear relations between integrals, most importantly the different integration-by-parts (IBP) routines, fail at finite temperature because of the
discrete nature of the zeroth component of the integration momenta. New ideas are warmly welcomed.

\subsection{The soft scales}

The four-loop calculations one needs to perform in the effective three-dimensional theories to obtain their contribution to the QCD pressure up
to order $g^6\ln\,g$ --- let alone $g^6$ --- are highly non-trivial already at the level of generating the diagrams and employing the IBP relations to
find the master integrals
(see Fig.~2). These computations have, however, been described in some detail already in Ref.~\cite{yorkll}, so we will restrict the present treatment to
cover merely the numerical evaluation of the four-loop masters performed in Ref.~\cite{avyork}. The strategy we chose to follow there was based on
deriving
linear difference equations in propagator powers for the integrals and solving them numerically via factorial series, which was proposed and developed by
S.~Laporta in Ref.~\cite{lapo}. Below we comment on our implementation of the algorithm only very briefly; for details,
see Refs.~\cite{avyork,lapo}.

%%%%%%%%%%%%%%%%%%%%%%%%%%%%%%
\begin{figure}[t]
\centerline{\epsfxsize=8.0cm \epsfbox{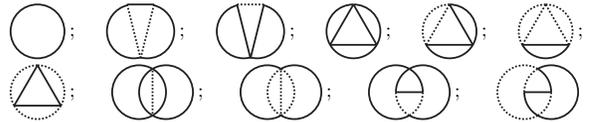}}
\caption[a]{The one- (1), two- (0), three- (2) and four-loop (8) scalar master diagrams of EQCD \cite{klry2}. The solid lines correspond to
massive and dotted to massless scalar propagators.}
\end{figure}
%%%%%%%%%%%%%%%%%%%%%%%%%%%%%%%%%%%%

Let $U$ be a generic master integral. The basic idea behind the difference equation approach is to attach an arbitrary power $x$ to one of its lines,
\ba
U(x)\;\equiv\;\int\fr{1}{D_1^xD_2...D_N},
\ea
and by employing IBP identities in a systematic way to derive for it a linear equation of the form
\ba
\sum_{j=0}^{R} p_{j}(x)U(x+j)&=&F(x), \label{deq}
\ea
where $R$ is a finite integer, the $p_j$'s polynomials in $x$ and $d$, and the right-hand side $F(x)$ a function of simpler (already known) integrals.
By substituting a factorial series ansatz into Eq.~(\ref{deq}) one can then derive recursion relations for the coefficients, and truncating the series at
some finite $s_{max}$ one finally obtains a result of desired accuracy for the initial graph at some high $x_{max}$. The last step is then simply to
push the solution back to $x=1$ by using the above difference equation, which however usually leads to a severe loss of accuracy in the results; hence a
careful optimization of the ratio $s_{max}/x_{max}$ is required.

In our computations we employed the FORM program \cite{form} to build the difference equations and recursion relations,
and finally Mathematica to obtain the numerical solutions and expand them in powers of $\e$. The first step utilized a
slightly modified version of the IBP algorithm described in Ref.~\cite{yorkll} to construct the equations and the
routines introduced in Ref.~\cite{lapo} to solve them in terms of factorial series. The Mathematica part on the
other hand consisted of numerically solving the recursion relations of the factorial series coefficients, performing the
actual summation and finally implementing the push-back step. We used several tricks to increase the speed of the
computations, some of which are explained in Ref.~\cite{avyork}. An essential feature in the project was to attack the graphs in a specific order from
the simplest one-loop case to the most
complicated four-looper, which ensured that the inhomogeneous terms of the difference equations were known at each step.

\section{SMALL $T/\mu$: EXPLICIT RESUMMATIONS}

To conclude, let us still briefly review the status of perturbation theory in the case of cold and dense quark matter, for which $T/\mu\ll 1$. It is
obvious that in this limit the framework of dimensional reduction must cease to work, as the
temperature no longer is the dominant energy scale in the problem. This can be explicitly verified in the exactly solvable large flavor number limit of
the theory \cite{avippreb}, where it is clearly seen that for values $T/\mu\leq 1/10$ the dimensionally reduced and exact results for the pressure start
to deviate rapidly. In this region one is therefore forced to tackle the infrared problems --- which again occur at three loops --- through
an explicit resummation of the IR divergent ring diagrams. These graphs are shown in Fig.~3, and one needs to add them to the sum of the known
one-, two- and three-loop gluon-skeletons (2PI w.r.t. gluon lines) in order to render the whole result finite.

%%%%%%%%%%%%%%%%%%%%%%%%%%%%%%
\begin{figure}[t]
\centerline{\epsfxsize=8.0cm \epsfbox{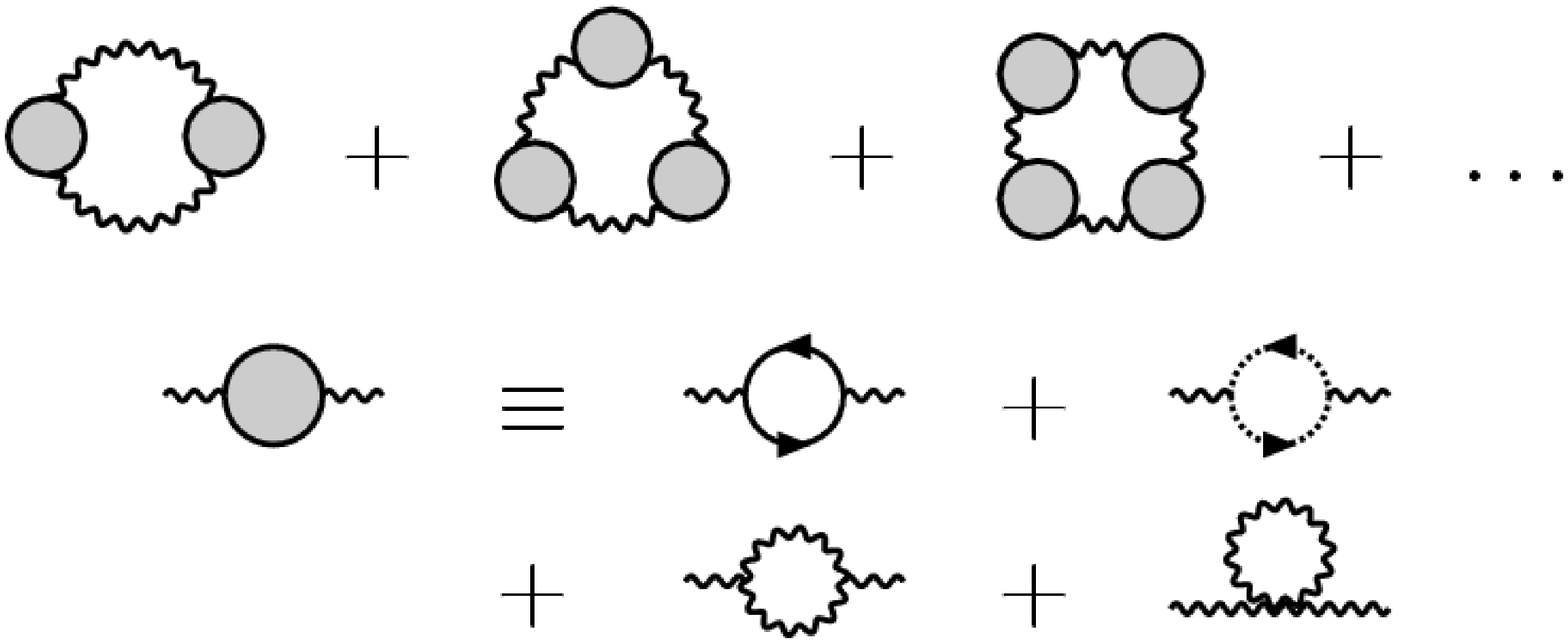}}
\caption[a]{The ring diagrams contributing to the QCD pressure at ${\mc O}(g^4)$.}
\end{figure}
%%%%%%%%%%%%%%%%%%%%%%%%%%%%%%%%%%%%

The evaluation of the ring sum at arbitrary $T$ and $\mu$ is a highly non-trivial problem even numerically. One simplification nevertheless occurs: since
the vacuum part of the one-loop gluon polarization tensor behaves as $\sim\! P^2$ in the IR limit, its contribution to the diagrams can be expanded as a
power series in $g^2$. This leads to a clean separation of all the singularities and renormalization scale dependence from the sum, and leaves an
entirely finite quantity to be evaluated numerically. This task is being attacked at present \cite{ikrv}.


\begin{thebibliography}{9}


%\cite{Kajantie:2002wa}
\bibitem{klry}
K.~Kajantie, M.~Laine, K.~Rummukainen and Y.~Schr\"oder,
%``The pressure of hot QCD up to g**6 ln(1/g),''
Phys.\ Rev.\ D {\bf 67} (2003) 105008 [hep-ph/0211321].
%%CITATION = HEP-PH 0211321;%%


%\cite{Laine:2003ay}
\bibitem{lainesewm}
M.~Laine,
%``What is the simplest effective approach to hot QCD thermodynamics?,''
Proceedings of SEWM 2002 [hep-ph/0301011].
%%CITATION = HEP-PH 0301011;%%

%\cite{Vuorinen:2002ue}
\bibitem{avsusc}
A.~Vuorinen,
%``Quark number susceptibilities of hot QCD up to g**6 ln(g),''
Phys.\ Rev.\ D {\bf 67} (2003) 074032
[hep-ph/0212283].
%%CITATION = HEP-PH 0212283;%%


%\cite{Vuorinen:2003fs}
\bibitem{avpres}
A.~Vuorinen,
%``The pressure of QCD at finite temperatures and chemical potentials,''
Phys.\ Rev.\ D {\bf 68} (2003) 054017
[hep-ph/0305183].
%%CITATION = HEP-PH 0305183;%%


\bibitem{ikrv}
A.~Ipp, K.~Kajantie, A.~Rebhan and A.~Vuorinen,
in progress.


%\cite{Braaten:1995jr}
\bibitem{bn1}
E.~Braaten and A.~Nieto,
%``Free Energy of QCD at High Temperature,''
Phys.\ Rev.\ D {\bf 53} (1996) 3421
[hep-ph/9510408].
%%CITATION = HEP-PH 9510408;%%



%\cite{Hart:2000ha}
\bibitem{hlp}
A.~Hart, M.~Laine and O.~Philipsen,
%``Static correlation lengths in QCD at high temperatures and finite  densities,''
Nucl.\ Phys.\ B {\bf 586} (2000) 443
[hep-ph/0004060].
%%CITATION = HEP-PH 0004060;%%


%\cite{Burgio:1997fp}
\bibitem{yorkpar}
F.~Di Renzo, A.~Mantovi, V.~Miccio and Y.~Schr\"oder,
%``Four loop stochastic perturbation theory in 3d SU(3),''
hep-lat/0309111.
%%CITATION = HEP-LAT 0309111;%%


%\cite{Linde:ts}
\bibitem{linde}
A.~D.~Linde,
%``Infrared Problem In Thermodynamics Of The Yang-Mills Gas,''
Phys.\ Lett.\ B {\bf 96} (1980) 289.
%%CITATION = PHLTA,B96,289;%%



%\cite{Arnold:ps}
\bibitem{az}
P.~Arnold and C.~X.~Zhai,
%``The Three Loop Free Energy For Pure Gauge QCD,''
Phys.\ Rev.\ D {\bf 50} (1994) 7603
[hep-ph/9408276];
%%CITATION = HEP-PH 9408276;%%
%\cite{Arnold:1994eb}
%``The Three loop free energy for high temperature QED and QCD with fermions,''
Phys.\ Rev.\ D {\bf 51} (1995) 1906
[hep-ph/9410360].
%%CITATION = HEP-PH 9410360;%%





%\cite{Gynther:2003za}
\bibitem{antti}
A.~Gynther,
%``The electroweak phase diagram at finite lepton number density,''
Phys.\ Rev.\ D {\bf 68} (2003) 016001
[hep-ph/0303019].
%%CITATION = HEP-PH 0303019;%%



\bibitem{avdis}
A.~Vuorinen,
%``The pressure of QCD at finite temperature and quark number density,''
hep-ph/0402242.
%%CITATION = HEP-PH 0402242;%%



%\cite{Schroder:2002re}
\bibitem{yorkll}
Y.~Schr\"oder,
%``Automatic reduction of four-loop bubbles,''
Nucl.\ Phys.\ Proc.\ Suppl.\  {\bf 116} (2003) 402
[hep-ph/0211288].
%%CITATION = HEP-PH 0211288;%%


\bibitem{avyork}
Y.~Schr\"oder and A.~Vuorinen,
%``High-precision evaluation of four-loop vacuum bubbles in three dimensions,''
hep-ph/0311323
%%CITATION = HEP-PH 0311323;%%



%\cite{Laporta:2001dd}
\bibitem{lapo}
S.~Laporta,
%``High-precision calculation of multi-loop Feynman integrals by  difference equations,''
Int.\ J.\ Mod.\ Phys.\ A {\bf 15} (2000) 5087
[hep-ph/0102033].
%%CITATION = HEP-PH 0102033;%%




%\cite{Kajantie:2003ax}
\bibitem{klry2}
K.~Kajantie, M.~Laine, K.~Rummukainen and Y.~Schr\"oder,
%``Four-loop vacuum energy density of the SU(N(c)) + adjoint Higgs theory,''
JHEP {\bf 0304} (2003) 036
[hep-ph/0304048].
%%CITATION = HEP-PH 0304048;%%



%\cite{Vermaseren:2000nd}
\bibitem{form}
J.~A.~Vermaseren,
``New features of FORM,''
math-ph/0010025.
%%CITATION = MATH-PH 0010025;%%


%\cite{Ipp:2003jy}
\bibitem{avippreb}
A.~Ipp and A.~Rebhan,
%``Thermodynamics of large-N(f) QCD at finite chemical potential,''
JHEP {\bf 0306} (2003) 032
[hep-ph/0305030].
%%CITATION = HEP-PH 0305030;%%
%\cite{Ipp:2003yz}
A.~Ipp, A.~Rebhan and A.~Vuorinen,
%``Perturbative QCD at non-zero chemical potential: Comparison with the
%large-N(f) limit and apparent convergence,''
Phys.\ Rev.\ D {\bf 69} (2004) 077901
[hep-ph/0311200].
%%CITATION = HEP-PH 0311200;%%





\end{thebibliography}
\end{document}